\documentclass{article}

\usepackage{arxiv}

\usepackage[utf8]{inputenc} 
\usepackage[T1]{fontenc}    
\usepackage{hyperref}       
\usepackage{url}            
\usepackage{booktabs}       
\usepackage{amsfonts}       
\usepackage{nicefrac}       
\usepackage{microtype}      
\usepackage{lipsum}
\usepackage{graphicx}
\usepackage{amsmath}
\usepackage{mhchem}
\graphicspath{ {./images/} }

\title{Leveraging Low Index Contrast to Reduce the Polarization Anisotropy in One-Dimensional Photonic Crystals}

\author{
 Jonathan Barolak \\
  Dipartimento di Fisica\\
  Università degli Studi di Pavia\\
  Via Bassi 6, 27100\\
  Pavia, Italy \\
  \texttt{jonathanjames.barolak@unipv.it} \\
   \And
 Agostino Occhicone \\
  Dipartimento di Scienze di Base e Applicate per l’Ingegneria\\
  Sapienza Università di Roma\\
  Via A. Scarpa 16, 00161\\
  Roma, Italy \\
  \And
 Marco Finazzi \\
  Dipartimento di Fisica\\
  Politecnico di Milano\\
  Piazza Leonardo Da Vinci 32, 20133\\
  Milano, Italy \\
    \And
 Paolo Biagioni \\
  Dipartimento di Fisica\\
  Politecnico di Milano\\
  Piazza Leonardo Da Vinci 32, 20133\\
  Milano, Italy \\
    \And
  Giovanni Pellegrini \\
  Dipartimento di Fisica\\
  Università degli Studi di Pavia\\
  Via Bassi 6, 27100\\
  Pavia, Italy \\
  \texttt{giovanni.pellegrini@unipv.it} \\
}

\begin{document}
\maketitle
\begin{abstract}
One-dimensional photonic crystals (1DPCs) are widely used optical platforms for guiding, filtering, and enhancing light at the nanoscale. Traditionally, designs have favored high refractive index contrast to maximize the size of the photonic band gap. Here, we demonstrate that low-index contrast systems offer a powerful and underexplored route to achieving improved optical functionalities. In particular, we show that low index contrast enables more closely aligned photonic band gaps for transverse electric (TE) and transverse magnetic (TM) polarizations, allowing for broadband superposition of TE and TM Bloch Surface Waves (BSWs). As a demonstration of this functionality, we use this approach to design 1DPCs capable of generating planar superchiral fields for enhanced circular dichroism spectroscopy. To realize such structures, we introduce an automated design framework based on multi-objective genetic optimization. By comparing optimized designs in both high and low index contrast regimes, we find that low index contrast systems yield significantly greater overlap between the TE and TM BSW dispersion relations and exhibit a pronounced reduction in optical anisotropy—both of which contribute to enhanced optical chirality across the operational bandwidth. Furthermore, numerical simulations reveal that these low index contrast structures offer improved robustness to fabrication tolerances and support a wider dynamic range of chiral analyte concentrations. In addition to their optical performance advantages, low-index contrast systems are naturally compatible with polymeric materials, which offer benefits such as low cost, environmental sustainability, and mechanical flexibility. While this work focuses on mode alignment and its consequences for surface wave behavior, the underlying design principle—enabled by low index contrast—has broader implications for polarization-independent photonic technologies, including optical sensing, computing, and spectral filtering.
\end{abstract}


\section*{Introduction}

Photonic crystals are foundational elements in nanophotonics, offering precise control over light propagation, confinement, and spectral/angular filtering through engineered periodic dielectric structures \cite{yablonovitchInhibitedSpontaneousEmission1987,johnStrongLocalizationPhotons1987,gangwarRecentProgressPhotonic2023}. The simplest version is the one dimensional photonic crystal (1DPC), also known as the photonic crystal multilayer, which has alternating low- and high- refractive index, planar materials. Although nomenclature varies across the literature, in this article we use the term 1DPC to refer to structures that may or may not exhibit periodicity in their layer thicknesses. One-dimensional photonic crystals (1DPCs) have found widespread use due to their simplicity and versatility in applications ranging from biosensing and waveguiding to topological photonics with uses in optical communications, optical switching, and photonic integrated circuits \cite{liControllingTETMSplitting2024,konopskyPhotonicCrystalSurface2007,descrovi2007}. In all application areas, the propagation of light through the 1DPC is governed by the devices photonic band structure, the optical analogy to the electronic band structure in crystalline semiconductors, and the associated photonic bandgap (PBG) \cite{joannopoulosPhotonicCrystalsMolding2008}. A strong trend in the photonic crystals community is to prefer 1DPC designs with large PBGs as they generally lead to better light control over a larger band of photon energies, have higher reflectivity coefficients, and stronger mode confinement \cite{aurelioElectromagneticFieldEnhancement2017a,finkDielectricOmnidirectionalReflector1998a,edringtonPolymerBasedPhotonicCrystals2001,mullerPhotonicCrystalFilms2000}. The trend is so strong that there is a substantial push in the community to increase the dielectric contrast of polymeric based systems to bring the utility of plastics, such as low production cost, ease to scale-up, elasticity, and potential to use recycled materials \cite{edringtonPolymerBasedPhotonicCrystals2001,lovaStrategiesDielectricContrast2020}, to photonic crystal devices. Though the design paradigm utilizing high index contrast devices has become a proven strategy for effective creation of 1DPC devices in applications such as reflectors, filters, and resonant optical structures, a new design challenge has arisen for 1DPC with requiring polarization independent behavior. 

Light entering a 1DPC can be decomposed into two polarization states: one with the electric field oscillating perpendicular to the plane of incidence (TE polarization), and one with the magnetic field oscillating perpendicular (TM polarization). The propagation of these two polarization states through a 1DPC is inherently different due to the optical anisotropy of the multilayer structure. The Brewster angle, where TM-polarized light is perfectly transmitted across an interface, is one prominent consequence of this anisotropy \cite{leeEssentialDifferencesTE2019,liControllingTETMSplitting2024}. Recently, however, there has been growing interest in polarization-independent filtering for optical computing, as well as polarization-independent surface state generation for superchiral light creation and multiplexed refractometric sensing using both polarization states \cite{xueHighNAOpticalEdge2021a,zhangIncoherentOptoelectronicDifferentiation2022,liuSinglePlanarPhotonic2022,pellegriniChiralSurfaceWaves2017,pellegriniSurfaceenhancedChiropticalSpectroscopy2018,pellegriniSuperchiralSurfaceWaves2019,mogniOneDimensionalPhotonicCrystal2022,pitruzzelloPhotonicCrystalResonances2018}. Although these applications can often operate in angular and spectral regions away from the Brewster angle—where closer alignment of the TE and TM PBGs may be possible — the design of such polarization-independent 1DPCs remains a challenge. While low-index contrast systems have generally been overlooked due to their weaker confinement and narrower PBGs, they represent a largely untapped design space with unique optical properties.


In this work, we demonstrate that low-index contrast 1DPCs offer a powerful and underexplored route to achieving novel optical functionalities through the close alignment of the TE and TM photonic band gaps. We explicitly show how utilizing such a design approach allows for broadband superposition of TE and TM Bloch Surface Waves (BSW) for the generation of planar superchiral light fields \cite{yehElectromagneticPropagationPeriodic1977}. Furthermore, we show how this design paradigm leads to 1DPC structures with reduced anisotropy in the field confinement leading to an enhanced optical chirality at the surface with an almost perfectly circular polarization state at variance with previous works that predicted a strong ellipticity with high index contrast 1DPCs \cite{pellegriniChiralSurfaceWaves2017,pellegriniSurfaceenhancedChiropticalSpectroscopy2018,pellegriniSuperchiralSurfaceWaves2019,mogniOneDimensionalPhotonicCrystal2022}. To realize these structures, we develop a multi-objective genetic optimization framework that identifies layer thicknesses which support TE–TM mode overlap across a targeted spectral range. We compare designs in both low and high index contrast regimes and show that only the low-index contrast systems achieve consistent modal alignment and reduced anisotropy. The low contrast designs result in improved generation of planar superchiral fields, enhanced robustness to fabrication tolerances, and a broader operating range for chiral analyte detection. Beyond chiral sensing, the underlying principle demonstrated here—polarization mode alignment through low index contrast design—has broader implications for polarization-independent photonic technologies. The compatibility of low-contrast designs with flexible, cost-effective, and environmentally friendly polymeric materials further expands their potential in scalable and adaptive photonic systems, including spectral filtering, optical information processing, and integrated photonic circuits.

\section*{Results and Discussion}

\begin{figure}[h]
  \centering
  \includegraphics[width=0.75\textwidth]{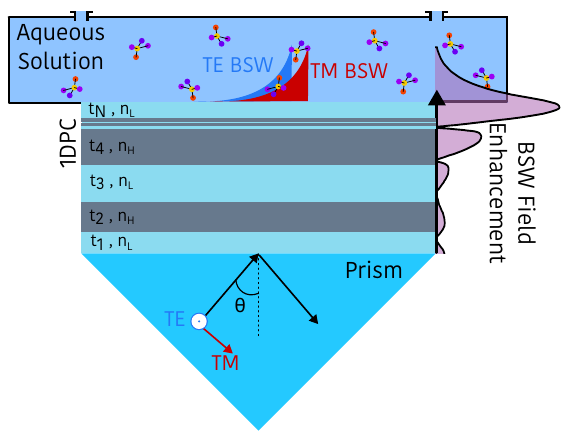}
  \caption{A diagram of a 1DPC for simultaneous excitation of TE and TM mode Bloch Surface Waves. The system is designed in the Kretschmann configuration with a BK7 prism as the substrate and an aqueous solution where the chiral analytes would exist as the superstrate.}
  \label{fig:diagram}
\end{figure}

Chirality is a fundamental property of matter in which an object's mirror image is not super imposable with itself despite being chemically identical. Chiral compounds are prolifically found in biological systems and pharmaceuticals where appropriately identifying the 'handedness' of the compound is crucial for safe and effective treatment \cite{laboratoryoftoxicologyfacultyofpharmacyuniversityofparis54avenuedelobservatoire75006parisfrance.ChiralDrugsOverview2006}. Optical diagnostic techniques, such as circular dichroism (CD) spectroscopy where the differential absorption between left and right circularly polarized light is measured, are commonly used for accurate identification of the handedness of chiral matter due to the relatively simple optical setup, fast data collection and processing times, and nondestructive nature. However, CD signals are typically much weaker than standard absorption spectroscopies thus posing a challenge for measuring small concentrations of chiral analytes. To increase the interaction of light with the chiral molecules, many solutions are coming forth to tailor the properties of the light field to increase their interaction with the chiral compounds as compared to plane wave illumination \cite{solomonNanophotonicPlatformsChiral2020a,biswasTunablePlasmonicSuperchiral2024,mohammadiNanophotonicEnhancedThermalCircular2025,munElectromagneticChiralityFundamentals2020,aliCircularDichroismPlasmonic2023}. Typically this is achieved by increasing the optical chirality, which can be defined as:

\begin{equation}
    C = -\frac{\epsilon_0 \omega}{2}Im(\textbf{E}^* \cdot \textbf{B})
    \label{eq:OpticalChirality}
\end{equation}

with $\epsilon_0$ being the free space permeability, $\omega$ being the optical angular frequency, and $Im()$ being the function that takes the imaginary component \cite{tangOpticalChiralityIts2010}. Circularly polarized light, for example, has an optical chirality of $C_{CPL} = \pm \epsilon_0 \omega /2$ and light with an optical chirality greater than this would be considered superchiral. Most nanophotonic solutions are based on the idea of increasing the field confinement and thus increasing the absolute value of $\textbf{E}$ and/or $\textbf{B}$ while maintaining a phase factor and polarization that sustains an elliptical, or ideally circular, polarized light state. It remains a challenge, however, to produce light with high optical chirality factors over a uniform surface, which can easily be switchable between left and right hand polarization, and which works from near-UV to IR wavelength ranges where chiral spectroscopy is critical. BSWs have recently been shown to generate planar superchiral light fields by confining light at the surface for both polarization states simultaneously. They operate by illuminating a truncated 1DPC at an angle and energy wavelength range such that the light cannot exit the top surface due to being in total internal reflection and the reflected light cannot propagate back through the crystal due to the presence of a PBG \cite{aurelioElectromagneticFieldEnhancement2017a}. A qualitative field profile of a BSW is depicted in figure \ref{fig:diagram} and simulated BSW field profiles are shown in figure \ref{fig:fields}.

An experimental system of the proposed superchiral field generation system is shown in figure \ref{fig:diagram}. The system uses the Kretschmann configuration to ensure that the incident light field is in total internal reflection on the top surface of the 1DPC \cite{occhiconeSpectralCharacterizationMidInfrared2021}. Light, containing both TE and TM polarization states with the appropriate phase offset, enters the 1DPC at an oblique angle from a prism made of N-BK7 glass. The fields propagate through the crystal and a strong field enhancement occurs at the top surface which is an aqueous solution where the chiral analytes would be suspended during CD spectroscopy. To generate the largest optical chirality at the surface, we want to generate a structure which contains BSW modes for both TE and TM polarization states at the same energy and momentum combination. To generate a large optical chirality over a range of wavelengths, which is required for CD spectroscopy, the dispersion relations of the TE and TM BSWs must be overlapped. Maximally overlapped dispersion relations occur when the band structures for each polarization state are aligned across the operational wavelength range of the CS spectroscopy system.


Overlapping TE and TM BSW modes is a challenging design criteria for 1DPCs. Previous attempts have aimed at utilizing a 'defect layer' at the termination of the 1DPC to align the dispersion relations of the BSWs, however this method came at the price of a strong anisotropy \cite{pellegriniChiralSurfaceWaves2017,pellegriniSuperchiralSurfaceWaves2019,pellegriniSurfaceenhancedChiropticalSpectroscopy2018,mogniOneDimensionalPhotonicCrystal2022}. Due to the anisotropy, the TM BSW was significantly weaker than the TE BSW, leading to a strong elliptical polarization states at the surface and a non-ideal optical chirality. To increase the optical chirality at the surface, by optimizing the overlap between the TE and TM BSW modes and reducing the anisotropy between the modes, we developed an automated design method using genetic optimization. 

\subsection*{Automated Design of 1DPC Structures} \label{section:AutoDesign}

Optimization of photonic crystal structures is a challenge due to the rough solution space and the large dimensionality of the problem especially when the constraint of a strict periodicity is lifted \cite{safdariMultiobjectiveOptimizationFramework2018,shiOptimizationMultilayerOptical2018}. While gradient based optimization methods struggle in these conditions, genetic algorithms, methods developed by mimicking the natural selection and genetic crossing/mutation in biological evolution, have been shown to excel. We use a genetic optimization method, specifically the Non-dominated Sorting Genetic Algorithm II (NSGA-II), to maximize the optical chirality generated at the top of our photonic crystal structure by optimizing the thicknesses of each layer within the 1DPC and the excitation angle of the light \cite{nsga2,DEAP_JMLR2012}. Our parameter vector is thus defined as:

\begin{equation}
    \vec{x}_{opt}: [t_1,t_2,...,t_N,\theta_1,\theta_2]
\end{equation}

\noindent where $t_i$ is the $i^th$ layer thickness and $\theta_{1,2}$ are the excitation angles for the wavelengths of light defining the operation wavelength range of the system. To properly align the dispersion relations of the TE and TM BSW modes over a range of wavelengths, a single objective function is not sufficient as the dispersion relations could cross in $(k_{\parallel},\omega$) space providing a 1DPC with a large optical chirality at only a single wavelength. For proper dispersion alignment of the BSW modes, we utilized a two-objective optimization method which aimed at overlapping the dispersion relations at the two extremes of the wavelength range of interest. Though the TE and TM BSW dispersion relations can have be characterized by a different second derivative, thus resulting in misaligned dispersion relations when the BSW modes are overlapped at the two extreme wavelengths, we are investigating a relatively small energy, momentum space where the dispersion relations are somewhat linear. Therefore, we can attempt to align the dispersion relations by writing an automated design method that aims to overlap the two modes at the wavelength extremes. The evaluation functions thus are defined as:

\begin{equation}
    Eval: [\frac{C_{\lambda_1}(z_s+z_0)}{C_{CPL}},\frac{C_{\lambda_2}(z_s+z_0)}{C_{CPL}}]
\end{equation}

\noindent where $C_{\lambda_1}$ is the optical chirality at wavelength $\lambda_1$, $z_s$ is the axial distance of the top surface of the photonic crystal, and $z_0$ is some offset distance to calculate the optical chirality within the aqueous solution where the chiral compounds would exist in the experimental setup. The optical chirality for each individual within the species are calculated by simulating the electric and magnetic fields within the proposed design from each individual. The simulated fields are calculated using a rigorous coupled wave analysis (RCWA) approach with only a single Fourier harmonic taken to force lateral isotropy within the layers \cite{whittakerScatteringmatrixTreatmentPatterned1999,vialOpenSourceComputationalPhotonics2022,nannos}. Plane wave solutions are assumed and the field is calculated assuming incident illumination at both wavelengths with their associated incident angles from the values in the parameter vector for the given individual within the population. Incident fields are assumed to have both polarization states present and the appropriate phase offset is calculated by maximizing the optical chirality at the distance $z_0$ above the top surface of the 1DPC.

As with all multi-objective optimization approaches, after the algorithm runs for many hundreds of iterations a single optimized result is not generated but rather a line of 'Pareto optimal' results forms in the ($Eval_1,Eval_2$) solution space \cite{censorParetoOptimalityMultiobjective1977}. A Pareto optimal individual represents an optimized design which cannot be improved in one of the evaluation functions without coming at the cost of the other evaluation function. The collection of Pareto optimal solutions forms a so called 'Pareto front'. An example Pareto front is shown in figure \ref{fig:opt_results}. In this manuscript, we chose the Pareto optimal solutions whose evaluation functions summed together are maximal as our optimal solution.

\subsection*{Comparison of High and Low Index Contrast Optimized Designs}

One of the free parameters we have while using our genetic optimization method is the choice of materials for the high and low refractive index. The material choice, and thus the index of refraction contrast, will have a strong impact on the photonic band structure and thus it makes sense to choose a refractive index contrast that will best align the photonic band structure for the two polarization states. We can gain a sense for how the index of refraction contrast affects the propagation between the two polarization states by using the ubiquitous transfer matrix method, which relates the forward and backward propagating waves at one layer of a 1DPC to the forward and backward propagating waves at another layer \cite{yehElectromagneticPropagationPeriodic1977}. This is done by assuming a plane wave solution to the electric field with amplitudes $E_{f,i}$ and $E_{b,i}$ for the forward and backward propagating waves in layer $i$ which can be expressed as the vector $\begin{pmatrix} E_{f,i} \\ E_{b,i} \end{pmatrix}$. By calculating the transfer matrix $M$ one can calculate the field in the next layer through: $ \begin{pmatrix} E_{f,i+1} \\ E_{b,i+1} \end{pmatrix} = \textbf{M} \cdot \begin{pmatrix} E_{f,i} \\ E_{b,i} \end{pmatrix}$. Due to linearity, this process can be repeated to calculate the field at any location within the crystal by multiplying together the transfer matrices of subsequent layers:

\begin{equation}
    M =  \prod\limits_i^I \textbf{M}_i = \begin{bmatrix} \prod\limits_i^I \textbf{B}_{i-1,i} \textbf{P}_i \end{bmatrix} \textbf{B}_{I,I+1}
\end{equation}

\noindent where $\textbf{B}_{i-1,i}$ is the boundary matrix between the $(i-1)^{th}$ and $i^{th}$ layer and $\textbf{P}_i$ is the propagation matrix within the $i^{th}$ layer \cite{grygaSensingBasedBloch2021}. While the propagation matrices don't depend on the polarization state, the boundary matrix does depend on the polarization state since the reflectance and transmittance through a discontinuity in the refractive index depends on the polarization state of the incident light. The boundary matrices are given by:

\begin{equation}
    \textbf{B}_{i,i+1} = \frac{1}{2}\begin{pmatrix} 1+\eta_{TE,TM} & 1-\eta_{TE,TM} \\ 1-\eta_{TE,TM} & 1+\eta_{TE,TM} \end{pmatrix}
\end{equation}

\noindent where the $\eta_{TE,TM}$ parameters can be related to each other through:

\begin{equation}
    \eta_{TM} = \frac{n_i^2}{n_{i+1}^2} \eta_{TE}.
    \label{eq:eta}
\end{equation}

Investigating equation \ref{eq:eta}, a natural choice in materials arises when trying to align dispersion relations between the TE and TM modes. Since the band diagrams are going to be intimately related to the equations relating the propagation of waves across the boundaries within the photonic crystal, it makes sense that we should choose materials that minimize the dissimilarity between the transfer matrices for the 1DPC of both polarization states. As one can see, the only difference in the transfer matrix for the two polarization states is the $\eta_{TE,TM}$ parameter in equation \ref{eq:eta} which is minimized when $n_i^2 = n_{i+1}^2$. Since the materials cannot be identical, as identical materials wouldn't create a PBG since the 1DPC would reduce to a simple slab of glass, by using materials with low-index contrast we would expect more similar transfer matrices for incident TE and TM polarization states. Though it is well known that the dispersion relations cannot be identical, due to the presence of Brewster's angle for only TM polarization states, we only need the dispersion relations to be closely aligned and not exactly aligned. This is due to the fact that our BSW are created on a finite layered 1DPC thus our BSW states will have a relatively low quality factor such that we can experimentally couple to the modes unlike many 1DPCs which can have tens of layers forming modes with extremely narrow resonances. 

\begin{figure}[h]
  \centering
  \includegraphics[width=1.0\textwidth]{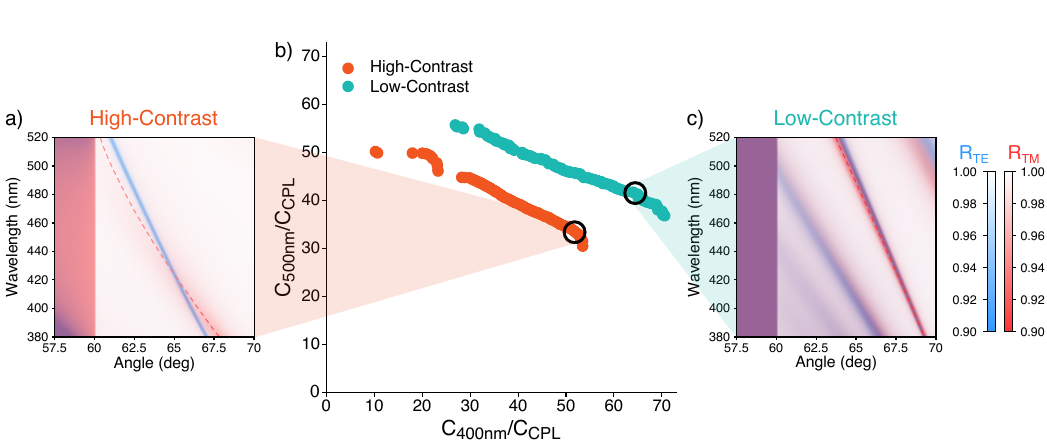}
  \caption{Pareto fronts from a multi-objective optimization run with both a high and low index contrast system are presented in figure b). In these optimizations, the thicknesses of a 1DPC were tailored to optimize the optical chirality enhancement at two different wavelengths. Reflectivity maps of one of the optimized designs from the high and low index contrast designs are presented in plots a) and c) with TE and TM modes being plotted in blue and red respectively. Dashed red lines are included to easily see the maximum of the TM BSW mode at each wavelength.}
  \label{fig:opt_results}
\end{figure}

To see whether the physical intuition of using a low index contrast design allows for improved overlap of the TE and TM BSWs, genetic optimization runs were performed using both design philosophies. To ensure a fair comparison, the optimization parameters were held constant including cross-over probability and clustering parameters. Each run went for 300 generations with 300 individuals in each generation and optimized the chirality at a position $z_0 = 5$ nm above the surface of the top layer for $\lambda_1 = 400$ nm and $\lambda_2 = 500$ nm. For the purposes of this analysis, our criteria for a fair comparison between the two optimized designs is whether the TE BSW modes have an equal width between the two designs, which is achieved with a different number of layers. Indeed, due to the inherent increase in the PBG size for the high-index contrast design, for an even number of layers the TE BSW will always results in much stronger field confinement. However, the resonance will be significantly narrower than the optimized low-index contrast BSW mode for the same number of layers. In these designs, we are examining 1DPCs with fewer than 15 layers such that the resulting BSWs have FWHM values that are experimentally viable for coupling. As such, it is always possible to add more layers to narrow the resonance and increase the field confinement. Therefore, it is logical to make the fair comparison between high and low index contrast designs given they have the same modal width instead of ensuring they have the same number of layers. 


Optimizations were performed for a 1DPC with $n_L = 1.50$ and $n_H = 1.60$ for the low index contrast design and $n_L = 1.50$ and $n_H = 2.55$ for the high index contrast design. The absorption value was held constant at $k = 0.0001$ for all layers excluding the water superstrate and the glass substrate which contained no absorption. An analysis on the effect of the absorption value on these results is presented in the SI. To utilize the optimization method for the purpose of determining which design paradigm is preferable, material dispersion was not taken into account. However, results between real world materials with dispersion taken into account are presented in the SI and the same findings are displayed in those optimizations. The high index contrast design was optimized with 5 layers of alternating low and high index material with the low index materials being the initial and final layers of the 1DPC. The low index contrast design had the same functional form but contained 13 layers. Fine tuning of the high contrast $n_H$ value was used to ensure the modal thickness were identical between the low and high index designs.

\begin{figure}[h]
  \centering
  \includegraphics[width=1.0\textwidth]{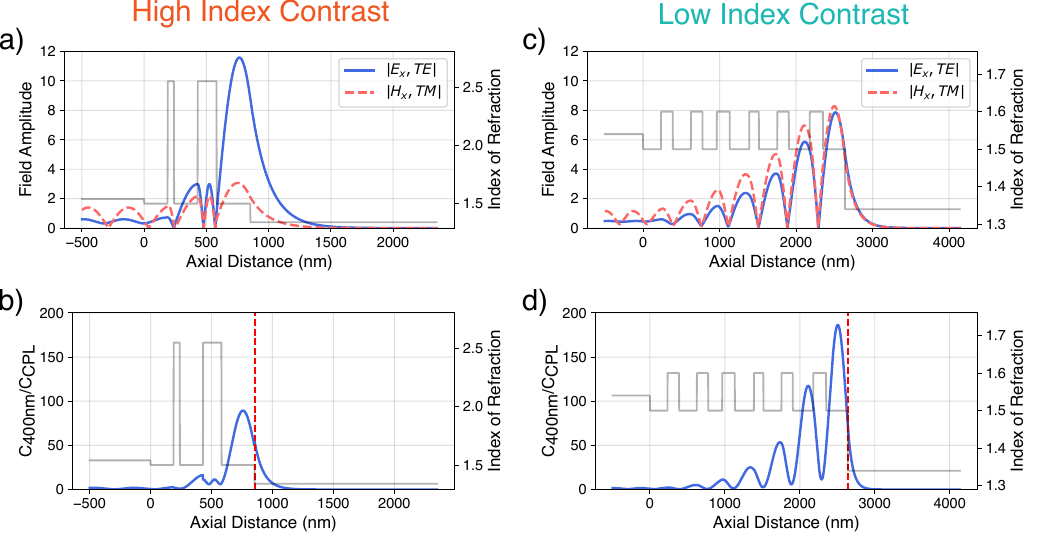}
  \caption{Axial field amplitude profiles, for only x polarization state, are shown for the high and low index contrast designs in plots a) and c) for a plane wave launched at the angle which optimizes the strength of the BSW for light with $\lambda = 400$ nm. Axial optical chirality enhancement profiles are presented in plots b) and d) with the red dashed line representing the axial point $z_0$ where the optical chirality is calculated for the genetic algorithm's evaluation functions.}
  \label{fig:fields}
\end{figure}

Results from the two-objective, genetic optimizations for both the high and low contrast design ideologies are presented in figure \ref{fig:opt_results}. The individuals plotted in ($C_{\lambda_1},C_{\lambda_2}$) space represent the individuals of the Pareto front of the last generation of the genetic optimization. The values of the optical chirality are normalized by the maximum optical chirality for CPL to show the optical chirality enhancement. The shapes of the Pareto fronts are similar however the low-contrast Pareto front fully dominates the high contrast Pareto front. This Pareto dominance indicates that there is an individual in the low contrast Pareto front which will offer an improvement in both chirality enhancement values for every individual in the high contrast Pareto front. To gain a better sense for what is causing this improvement, we can plot the reflectivity maps in ($\lambda$,$\theta$) space for the individuals in each Pareto front which maximize the sum of the individual's optical chirality enhancement values at the two wavelength extremes \cite{wankerlParameterizedReinforcementLearning2021}. In both reflectivity maps, the red hue represents the reflectivity of the TE mode and the blue hue represents the reflectivity of the TE mode. Due to the broad BSW resonance of the TM mode for the high-contrast design, a dashed line is plotted to show the angle of the minimum reflectivity dip for each wavelength. The reflectivity maps show us that the decreased optical chirality is caused by a decrease in the modal overlap of the high contrast design and a reduced anisotropy of the low contrast design. Specifically from the dashed lines, it is clear that the TE and TM BSW dispersion relations are not aligned well in the high-contrast case leading to a walk off of the BSW modes. In the low contrast case, the reflectivity maps show that the BSW modes for the TE and TM polarizations are almost fully overlapped. Additionally, we see a significant decrease in anisotropy in the low contrast design due to narrow TM BSW mode as compared to the broad TM BSW mode in the high contrast design. These results show that using low index contrast materials can lead to alignment of the TE and TM BSW modes which can in turn create a favorable optical chirality enhancement.

\begin{figure}[h]
  \centering
  \includegraphics[width=0.8\textwidth]{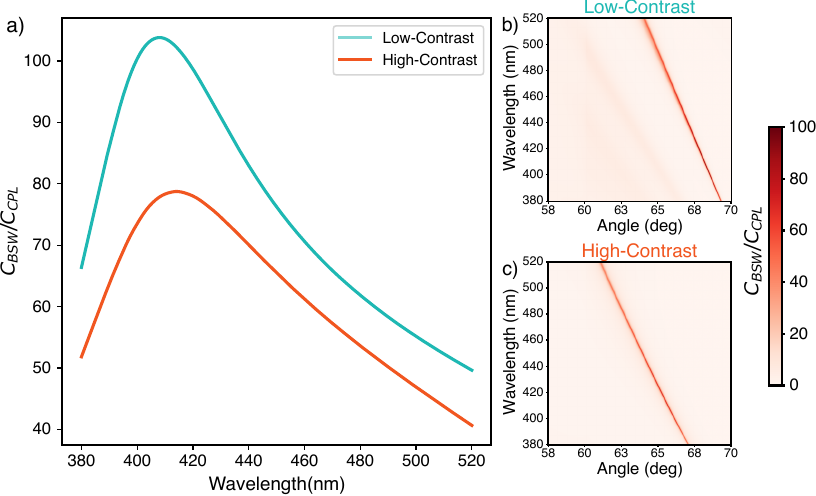}
  \caption{The optical chirality enhancement value is plotted in a) for both the high and low contrast designs over the entire operational bandwidth. The associated chirality maps are presented in b) and c).}
  \label{fig:chirality_maps}
\end{figure}

In the reflectivity map for the low contrast optimized design in figure \ref{fig:opt_results}, we see many modes to either sides of the BSW mode. These modes are guided modes and indicate the significantly smaller PBG as compared to the high contrast design. To ensure that we are looking at the BSW mode, we can plot the field amplitude, $|E_x(z)|$ and $|H_x(z)|$, and optical chirality enhancement, $C_{\lambda_1}(z)/C_{CPL}$, as a function of axial distance, $z$, through the 1DPC. The fields are calculated at $\lambda = 400$ nm and $\theta_{inc} = \theta_{400}$ where $\theta_{400}$ is the angle which maximizes the optical chirality at $400$ nm for each design respectively. The $x$ components of the electric and magnetic fields for the optimized high and low index contrast designs are plotted in figure \ref{fig:fields}. Here the fields are normalized to the amplitude of the incoming field. The $y$ and $z$ components of the fields are plotted in the SI. The index profiles are shown in gray in the background. In the substrate, we see the interference between the forward and backward propagating plane waves giving rise to a stationary field. Tracking the field amplitude of the low contrast design, we see an exponential increase of the field local maxima as the waves propagate through the crystal followed by the expected exponential decay in the superstrate. This behavior shows that the investigated mode in the reflectivity map is indeed a BSW and not a guided mode. Additionally, in the field profiles, it is clear to see the reduced anisotropy in the low contrast design. Comparing the field of the TE and TM modes in the superstrate, we see almost identical enhancements between the two modes showing that the surface wave has an almost perfectly circular polarization (when phased delayed by $\pi/2$ with respect to one another) while the high contrast design has strong ellipticity. The chirality enhancement as a function of axial distance are also plotted in figure \ref{fig:fields}. It is interesting to note that the ratio of the integrated energy outside the crystal divided by the integrated energy inside the crystal is larger for the high contrast case. This suggests that high contrast designs could be more sensitive to changes in the index of the superstrate in the frame of refractometric sensing. However superchiral sensing is based on absorption CD and the optical chirality enhancement factor at the location of the chiral analytes is the quantity used to determine how strongly the field will interact with a chiral molecule.

To assess the optical chirality performance throughout the entire operation wavelength range, we can plot the map of the optical criticality in ($\lambda$,$\theta_{inc}$) space. The resulting chirality maps, calculated at $z = z_0$ above the surface of the 1DPC, are shown in figure \ref{fig:chirality_maps}. Here the chirality values are calculated with the optimized phase offset for $\lambda$ and $\theta_{inc}$ within the map. We see in the low contrast chirality map, the guided modes have significantly lower chirality as compared to the BSW mode since their field distributions have localized field enhancement contained within layers of the 1DPC instead of at the surface. The chirality maps show that the superchiral light is only generated when the TE and TM BSW modes are overlapped. Running along the dispersion relations of both chirality maps, we can plot the maximum optical chirality as a function of angle for each wavelength. Plotted in figure \ref{fig:chirality_maps} we see the optical chirality at every wavelength is improved for the low contrast design as compared to the high contrast design. These results show that the low contrast design paradigm has superior performance to the high contrast designs for superimposed TE and TM BSW modes for generating planar superchiral light fields. 

\subsection*{Modal Width Analysis}

\begin{figure}[h]
  \centering
  \includegraphics[width=0.7\textwidth]{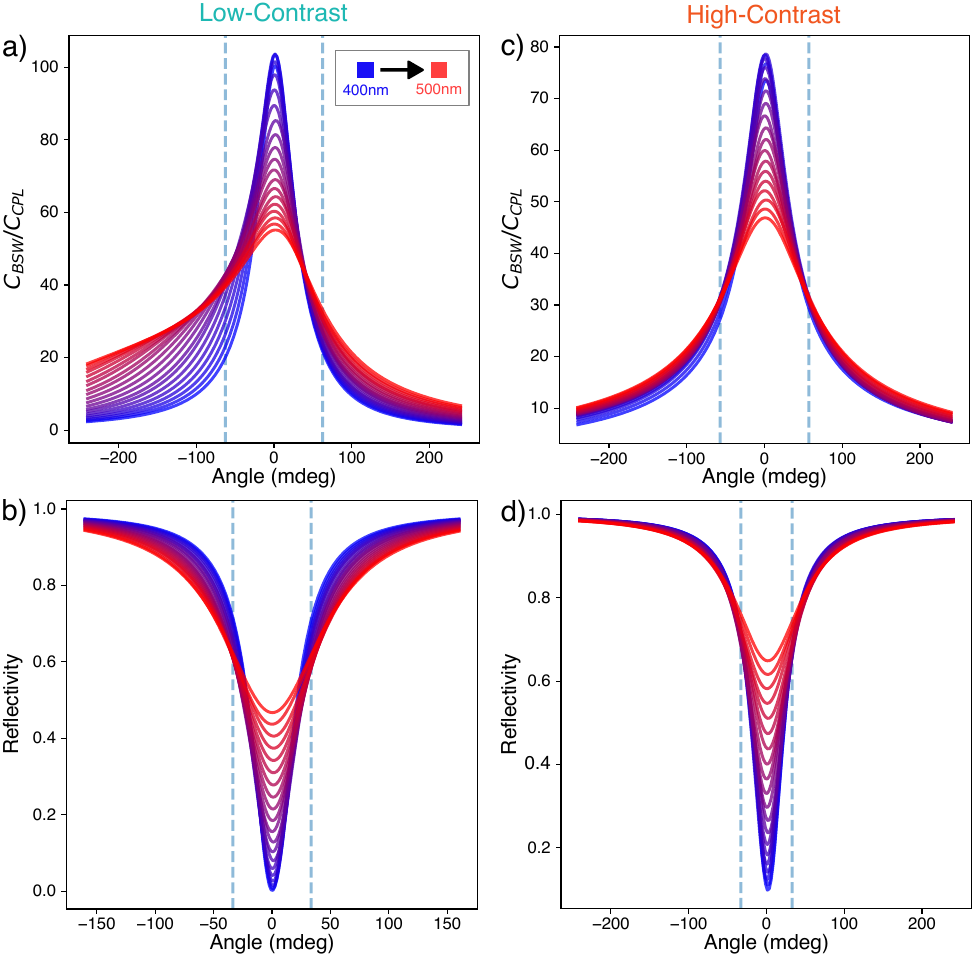}
  \caption{The modal widths for the high and low index contrast designs are presented for both the reflectivity maps b) and d) as well as the optical chirality enhancement maps a) and c). The spectra are calculated for many wavelengths between 400 nm and 500 nm, which are colored accordingly from blue to red. Vertical dashed lines are included to show the average FWHM calculated over all lineouts for a given plot.}
  \label{fig:modal_analysis}
\end{figure}

To ensure explicitly that we are analyzing fair designs, we can analyze the modal widths of the BSW for each design, presented in figure \ref{fig:modal_analysis}. To assess the modal widths, the reflectivity spectra of the TE BSW is plotted for various horizontal lineouts of the reflectivity map shown in figure \ref{fig:opt_results} for several wavelengths over the operational wavelength range. The minima have been centered for ease of comparison. The angular reflectivity spectra are colored on a linear color map which varies from blue to red as the wavelength ranges from $400$ nm to $500$ nm. Dashed lines are included to show the average FWHM over the entire wavelength range. The angular FWHM for the low contrast TE reflectivity dip is $67$ mdeg and the FWHM for the high contrast TE reflectivity dip is $66$ mdeg indicating that the comparison made in the previous section was indeed a fair comparison. The width of the TE BSW mode for the reflectivity dip was chosen as the criteria for fair comparison over the width for the chirality enhancement because the chirality enhancements take into account both the modal widths of the TE and TM BSWs as well as their relative overlap. Thus the TE BSW mode width is a better metric to use for the narrowness of the resonance and an indication of the confinement of the electric field. However, it is still useful to calculate the width of the chiral enhancement as this is the angular measurement required to know what resonance one can experimental couple into. Here we again see similar FWHM for the two designs with the low contrast having a FWHM of $120$ mdeg and the high contrast having a FWHM of $110$ mdeg. For an experimental system with improved angular uncertainty, capable of coupling to even narrower modes, one would expect to see an increased confinement and therefore an increased optical chirality enhancement. 

\subsection*{Stability Analysis}

\begin{figure}[h]
  \centering
  \includegraphics[width=1.0\textwidth]{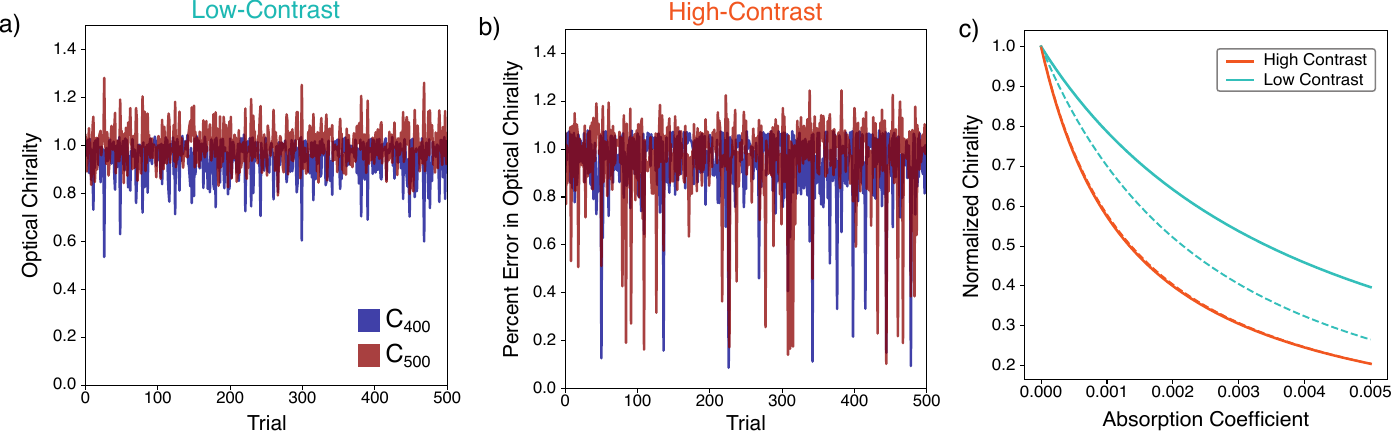}
  \caption{Plots a) and b) show the results of a Monte Carlo sensitivity analysis where many iterations of the 1DPC are generated with the thickness pulled from a random normal distribution with a standard deviation of 5 nm to simulate imperfections of the thickness in the manufacturing process. The optical chirality from each trial was calculated at both 400 nm and 500 nm illumination wavelengths. The plot in c) shows the normalized optical chirality enhancement as a function of absorption coefficient in the aqueous superstrate.}
  \label{fig:stability_analysis}
\end{figure}

In this section the effect of two forms of experimental instability are analyzed for the optimized 1DPCs from the high and low index contrast paradigms. The first source of instability is the common issue of instability generated from manufacturing errors. Though fabrication methodologies of 1DPCs have dramatically improved in recent decades, thickness errors are still a prevalent problem in the manufacturing of high-Q 1DPCs. To investigate our designs stability to manufacturing errors in the thickness of the 1DPCs, Monte Carlo Uncertainty Analysis was performed on the optimized designs. To simulate the manufacturing defects, a 1DPC was simulated with thicknesses chosen from a random normal distribution:

\begin{equation}
    t_i \sim \mathcal{N}(\mu_i, \sigma^2) \quad \text{for } i = 1, \ldots, N
\end{equation}

where $t_i$ is the thickness of the $i^{th}$ layer, $\mathcal{N}$ is the normal distribution function, $\mu_i$ is the actual thickness of the optimized design, and $\sigma$ is the standard deviation which was set to $5$ nm. The optical chirality angular spectrum was calculated for the faulty design at both design wavelengths and the maximum chirality enhancement values were recorded. This process is repeated for many hundreds of iterations to generate a distribution of optical chirality values for statistical analysis of the effect of the manufacturing errors. A dataset of the faulty chirality enhancement values, normalized to the 'ideal' case, are presented in figure \ref{fig:stability_analysis} a) and b) for the low and high index contrast optimized designs. Here the optical chirality enhancement of the individual with the manufacturing defects at 400 nm are plotted in blue and at 500 nm are plotted in red. It can easily be seen that the low index contrast designs are sizably more robust to errors in the thickness despite the fact that there are more layers for there to be errors within. The standard deviation for the normalize optical chirality enhancement is around $8 \%$ for the low contrast case and around $20 \%$ for the high contrast case. It may look surprising that there are enhancements to the optical chirality in both the low contrast and high contrast designs when the thicknesses are sampled from a random distribution centered around the optimized thicknesses. However, due to the multi-objective nature of the optimization approach, an enhancement in one design criterion can generally be made at the cost of the other criterion. A deeper analysis of this subject is presented in the SI. This analysis shows that the low contrast design paradigm also leads to designs which are significantly more robust to manufacturing errors than the standard high contrast designs. 

The second source of experimental instability is from absorption of the chiral analytes. The goal of generating superchiral light is to enhance the interaction of light with chiral media, thus lowering the required concentration in order to measure a detectable CD signal. Most of these techniques, including the 1DPCs, however work with relatively narrow resonances to generate the superchiral light which often require specific operating conditions for the resonance to be present. Absorption of the light in the region where the chiral analytes exist, for example, can ruin the resonance conditions. Thus these techniques typically bring with them a relatively small operating range of chiral analytes concentrations under which the superchiral light will exist. We can mimic this effect numerically by adding an absorbing layer on the top of our 1DPC sensor with a thickness of 1000 nm to act as our layer of chiral analytes. The index of refraction of this layer will be identical to water however the imaginary part of the refractive index will vary between $0.000 \leq k \leq 0.005$. At each absorption value, the optical chirality enhancement will be measured at the same point $z_0 = 5$ nm above the crystal, within the absorbing layer, as before. This was calculated using both the low index contrast and high index contrast design at both 400 nm and 500 nm. The chirality enhancement value, normalized to the case with no absorption, for each absorption value is displayed in the plot in figure \ref{fig:stability_analysis} c). Here the dashed line represented the normalized optical chirality enhancement for 500 nm while the solid line represents the value for 400 nm illumination. We see that the optical chirality decreases significantly faster for the high index contrast case than for the low index contrast case. For 500 nm illumination, the high index contrast has reached $50\%$ of its ideal optical chirality value for an absorption of 0.001 while the low index contrast design doesn't reach this level until an absorption value of over 0.003. This analysis shows that the low index contrast design also has a larger operational chiral analyte concentration range than the high index contrast design.

\section*{Conclusion}

In this research article, we showed how a low index contrast design approach can be used to achieve improved optical functionalities. Specifically we showed how a 1DPC with a low index contrast design can be used to significantly improve the alignment of the band structure and reflectivity maps of TE and TM light waves in a truncated 1DPC. To show the utility of such a design paradigm, we presented a multi-objective genetic optimization algorithm for generating optimized 1DPCs with tailored thicknesses for producing superchiral light fields, a challenging inverse design problem which requires alignment of TE and TM BSW modes. The optimized designs were then compared to show that the low index contrast design generated designs with improved optical chirality over the entire wavelength range of interest. By investigating the fields within the optimized crystals and the reflectivity spectra, we saw that the low contrast designs had significantly improved overlap of the TE and TM BSW modes and a large reduction in the anisotropy of the fields confined to the surface. Through Monte Carlo stability analysis and numerical simulations of the absorption of the confined fields at the surface we showed how the low contrast designs were significantly more robust to manufacturing errors in the thicknesses of each layer and were capable of operating with a larger range of chiral analyte concentrations on top of the 1DPC. The results from this analysis not only represents a major improvement in generating experimentally viable planar superchiral fields for sensitive chiral spectroscopy techniques, but also presents an important new design paradigm for polarization-insensitive 1DPCs which could have major impacts in fields like optical computing, sensing, and filtering. 

\section*{Acknowledgments}

This work was partially funded by the European Union-Next Generation EU-PNRR-M4C2, investimento 1.1-“Fondo PRIN 2022”-“SPIRAL – Lossless surface waves for chiral spectroscopy” – id 2022WFM5MZ – CUP F53D23001140001.

\section*{Supplemental Information}

This section contains the supplemental information for the above manuscript.

\subsection*{Optimized 1DPC Designs}

The Pareto fronts included in figure 2 of the main manuscript represent the optimized designs from the final generation that cannot be improved in one evaluation function without coming at the cost of the other evaluation function. These optimization runs optimized both the thicknesses of each layer of the 1DPC (between 30nm and 300nm) and the incident angle of excitation (constrained to be between 60 and 75 degrees) to optimize the optical chirality enhancement at both 400 nm and 500 nm illumination wavelength. The individuals with the largest summed evaluation functions from each Pareto front were chosen as the optimized design for investigation of various properties. 

\begin{table}[h]
\centering
\begin{tabular}{@{}ccc@{}}
\toprule
\textbf{Layer} & \textbf{High Index Contrast (nm)} & \textbf{Low Index Contrast (nm)} \\
\midrule
1   & Substrate     & Substrate     \\
2   & 189.8        & 235.7        \\
3   &  51.1        & 154.0        \\
4   & 189.9        & 237.0        \\
5   & 152.3        & 150.3        \\
6   & 270.6        & 187.4        \\
7   & Superstrate   & 168.7        \\
8   &               & 255.7        \\
9   &               & 125.5        \\
10  &               & 241.3        \\
11  &               & 151.1        \\
12  &               & 273.1        \\
13  &               & 171.5        \\
14  &               & 287.9        \\
15  &               & Superstrate   \\
\bottomrule
\end{tabular}
\caption{Layer thicknesses for the high and low index contrast 1DPC designs. Thicknesses are in nanometers.}
\label{tab:layer_thicknesses}
\end{table}

The optimized thicknesses of each optimized individual are presented in Table \ref{tab:layer_thicknesses} for completeness. The optimized angles for the high contrast design are $\theta_{400 nm} = 66.23$ and $\theta_{500 nm} = 62.13$ degrees. The optimized angles for the low contrast design are $\theta_{400 nm} = 68.56$ and $\theta_{500 nm} = 64.95$ degrees. The high contrast design had $n_H = 2.55$ for the odd numbered layers and $n_L = 1.50$ for the even numbered layers. The low contrast design had $n_H = 1.60$ for the odd numbered layers and $n_l = 1.50$ for the even numbered layers. The substrate for both designs was set to $n_{sub} = 1.54$ and the superstrate was set to $n_{sup} = 1.34$. The absorption value for all layers was $k = 0.0001$ except for the super and substrate which had no absorption.

\subsection*{Full Electric and Magnetic Field Plots}

\begin{figure}[h!]
  \centering
  \includegraphics[width=1.0\textwidth]{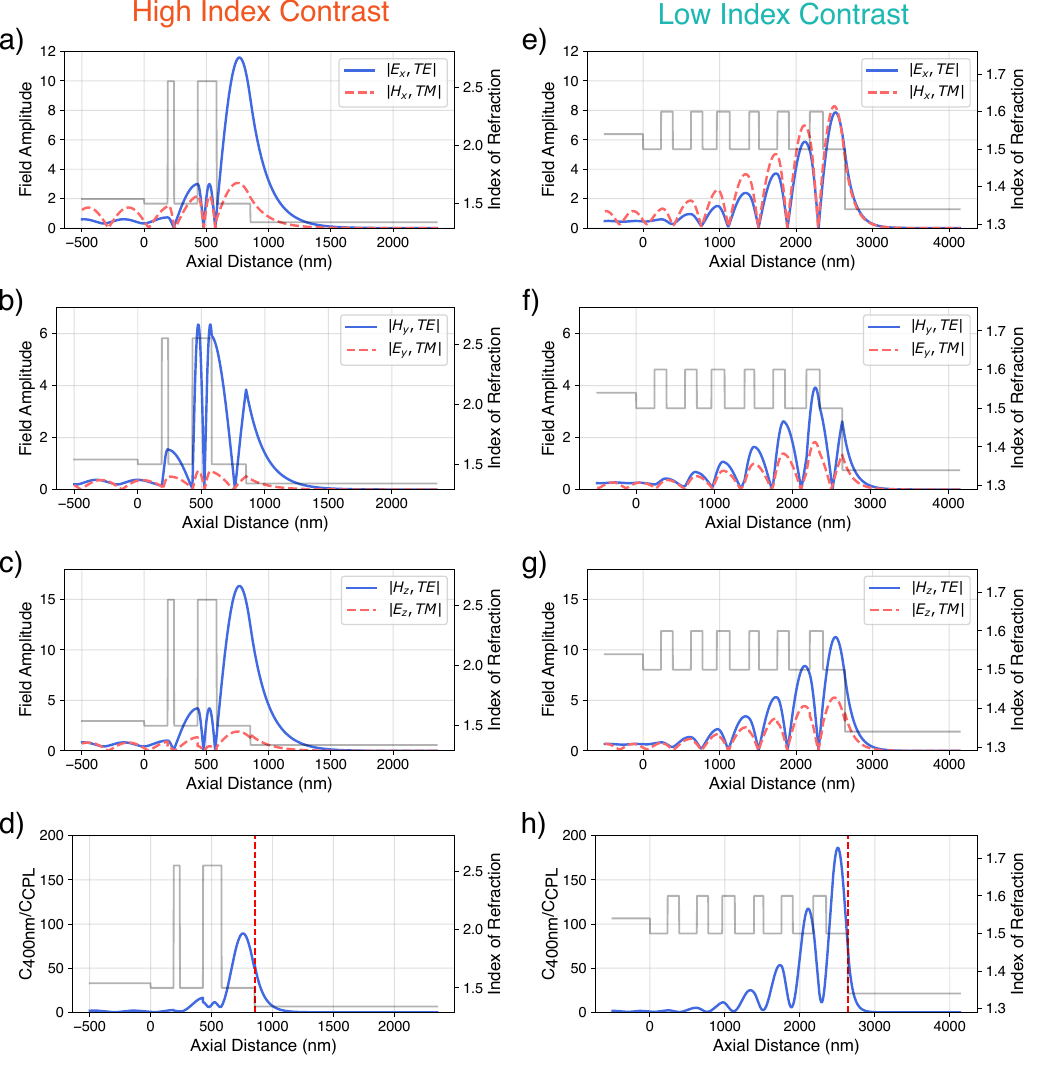}
  \caption{a) - c) show the x,y,z electric and magnetic field amplitude components normalized to the input electric field for the high index contrast design. e) - g) show the same plots for the low index contrast design. The TE mode is plotted with a solid blue line while the TM mode is plotted with a dashed red line. The gray plot in each image shows the axial modulation of the refractive index. d) and h) show the axial optical chirality enhancement for the high and low index contrast design respectively. }
  \label{fig:fields}
\end{figure}

In the main manuscript, figure 3 contained the axial field amplitude plots and chiral enhancement plots for the high and low contrast optimized designs as a function of the axial distance z. To constrain space and get across the message, only the x field component was included. Here we include all polarization directions for completeness. Figure \ref{fig:fields} reports the axial field amplitude distribution for all polarization directions for a plane wave launched with a wavelength of $\lambda = 400$ nm and an incident angle of $\theta_{400}$ presented in the \textit{Optimized 1DPC Designs} section of the SI. The field profiles are all normalized to the input electric field amplitude. Prior to reflection from the 1DPC, the field amplitude undulates according to the standing wave generated from the forward and backward waves generated from the excitation wave and the reflected wave. In the superstrate, the field decays exponentially, as expected, due to the total internal reflection condition. In the field plots, the TE fields are plotted with a solid blue line and the TM fields are plotted with a dashed red line. In all polarization directions, a reduced anisotropy between the TE and TM modes is present at the surface for the low index contrast design. The fields were calculated using the \textit{nannos} RCWA package and the thicknesses presented in the \textit{Optimized 1DPC Design} section of the SI \cite{nannos}.

\subsection*{Monte Carlo Uncertainty Analysis and Pareto Stability}

Investigation of the modal stability to manufacturing defects in the thicknesses of each layer was performed using a Monte Carlo Uncertainty Analysis. Here hundreds of 1DPCs were simulated with the thickness of each layer being taken as a random value on the normal distribution centered around the ideal thickness with a standard deviation of 5 nm. The optical chirality enhancement values were calculated at both 400 nm and 500 nm. This dataset was generated for both the high index contrast and low index contrast designs. Results, shown in figure 6 of the main manuscript, show that for one of the two evaluation functions the normalized optical chirality enhancement of several trials exceeds that of the ideal system. Given the multi-objective nature of the optimization, this makes sense as the random thicknesses can allow for movement of the individual along the Pareto front, thus increasing one evaluation function while coming at the cost of the other one. However, it is well known that genetic optimizations do not guarantee a converged solution, and therefore it is important to analyze whether adding randomness to the optimized thicknesses can simultaneously increase both objective values thus indicating that the optimizer had not fully converged. 

\begin{figure}[h]
  \centering
  \includegraphics[width=1.0\textwidth]{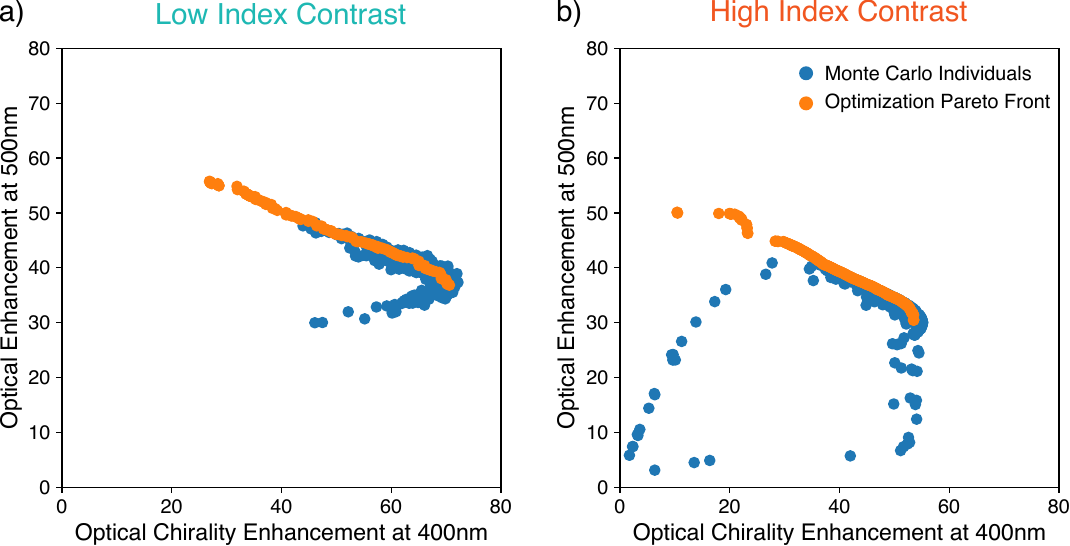}
  \caption{Pareto front analysis of the Monte Carlo uncertainty analysis simulation. In a) both the optimized Pareto front from the low index contrast systems, shown in orange, and the individuals from the Monte Carlo uncertainty analysis, shown in blue, are presented. Plot b) shows the results for the high index contrast systems.}
  \label{fig:Monte_Carlo_Pareto}
\end{figure}

The individuals generated from the Monte Carlo analysis in the $(C_{400},C_{500})$ space along with the optimized Pareto fronts are shown in figure \ref{fig:Monte_Carlo_Pareto} a) and b) for the low and high index contrast designs respectively. In panel a) we see that a subset of the individuals from the Monte Carlo dataset, plotted in blue, do in fact dominate the Pareto front from the multi-objective optimization for the low index contrast system. This indicates that the optimization likely had not fully converged and that more generations with more individuals would likely expand the Pareto front slightly beyond what was found given the optimization run which ran for 300 generations of 300 individuals. In b) however, we see almost no individuals from the Monte Carlo dataset dominate the Pareto front from the optimization with the high index contrast system indicating that this system had likely fully converged and adding more generations with more individuals would not likely improve the Pareto front greatly. This makes sense given that the low index contrast system has more parameters over which it needs to optimize due to the increase in the number layers to achieve the same modal width as the high index contrast system. Accordingly, the difference in performance between the two designs reported in this manuscript likely underrepresents the true performance gain from the low-index contrast system, which, due to its larger design space, would benefit more significantly from additional generations.


It is worth noting that in a) and b) of figure \ref{fig:Monte_Carlo_Pareto}, the individuals from the Monte Carlo dataset are centered about the selected design from the original Pareto front on which they are based. We see that exploration of these individuals due to random small thickness fluctuations expand the Pareto front slightly and fill out along directions which were not present in the original Pareto Front. More individuals and an adjusted crowding parameter would likely have found these solutions in the original optimization runs. Like the original plot of the stability in figure 6 of the main manuscript, in these plots we see that the high contrast system has individuals in the Monte Carlo dataset which are less clustered than the low contrast system, showing that the low index contrast design is more resilient to errors in the layer thicknesses of the optimized 1DPC.

\subsection*{Effect of Absorption within the 1DPC}

The performance difference between an optimized design with a high index contrast set of materials and a low index contrast set of materials will depend on the absorption value of the materials. Since the absorption value can vary greatly on fabrication conditions, temperature, material purity and defects, for fair comparison to address only the effect of material index contrast, the absorption was kept constant for all materials in the main manuscript. Here, we investigate how the value of the absorption coefficient, $k$ in $\tilde{n} = n + i k$, affects the results from the two color, optical chirality enhancement optimization. To compare the high and low contrast optimized designs for a variety of absorption values, optimization runs were performed and the number of layers in the low index contrast system were varied, while the high contrast system remained with 5 layers, until the width of the TE BSW mode between the high and low index contrast optimized designs were as close to equal as possible. This process was repeated for absorption values of $k=10^{-3}$ to $k=10^{-5}$ with the high index contrast design having refractive indices of $n_L = 1.50$ and $n_H = 2.55$ while the low index contrast system has $n_L = 1.50$ and $n_H =  1.60$. 

\begin{figure}[h]
  \centering
  \includegraphics[width=1.0\textwidth]{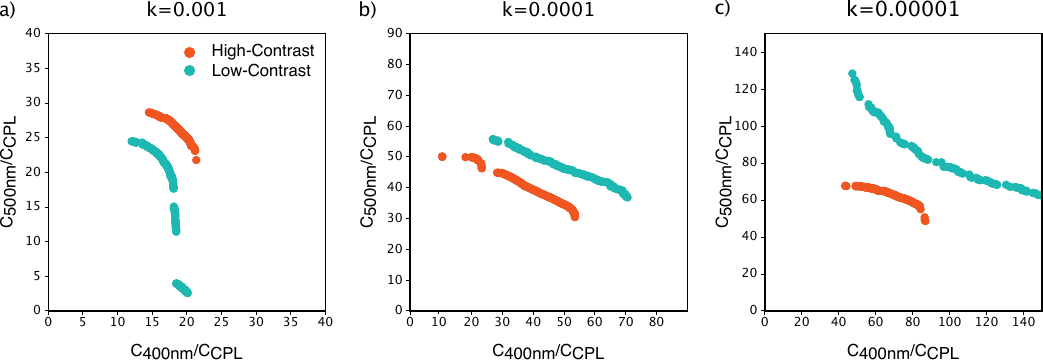}
  \caption{a) - c) show Pareto fronts for both the high and low index contrast systems for varying absorption avlues (imaginary part of the refractive index).}
  \label{fig:absorption_fig}
\end{figure}

Resulting Pareto fronts from each optimization run are shown in figure \ref{fig:absorption_fig}. It can be seen that the difference in Pareto fronts is drastically different between the 3 absorption values in a)-c). For the low absorption values, the low index contrast Pareto front shows a significant improvement to the high index contrast Pareto front. The improvement is smaller for the higher absorption value of $k=10^{-4}$ and in the case of $k=10^{-3}$ the high index contrast system actually outperforms the low index contrast system. The strong dependence of the performance difference between the two designs on the absorption coefficient value is likely due to the required difference in the number of layers between the two materials. We see that the low-index contrast systems are more affected by the value of the absorption coefficient because there are more layers in those designs due to the increased propagation distance within the materials. Absorption values lower that $k=10^{-5}$ do not have a strong effect on the performance of the sensors due to the fact that the fields have to propagate more than a millimeter in order for their intensity to be absorbed by $50\%$, calculated with $d=-\lambda / (4 \pi k) \ln{I/I_0}$, for optical wavelengths. For high quality optical materials, we would thus expect to see a strong performance improvement in terms of optical chirality enhancement for the low index contrast systems. 

\subsection*{Optimization with Real Materials}

The simulated 1DPC structures in the main manuscript were generated without taking into account material dispersion over the bandwidth of interest. Here we implement the same optimization routine but use index of refraction data from real materials to compose the layers of the high and low index contrast designs. The materials used for the high index contrast systems were \ce{SiO2} and \ce{Ta2O5} while the materials for the low index contrast systems were Cellulose and Polystyrene \cite{gaoExploitationMultipleIncidences2012,sultanovaDispersionPropertiesOptical2009,zhangComplexRefractiveIndices2020,10.1117/12.2227580}. The datasets were interpolated onto the wavelengths we used using a linear interpolation. These materials were chosen due to their prevalence in 1DPC designs for inorganic materials and polymeric systems \cite{lovaStrategiesDielectricContrast2020,mogniOneDimensionalPhotonicCrystal2022}. The absorption value was $k=0.00001$ for all layers except the substrate and superstrate which both had no absorption. To match the TE BSW modal widths between the optimized designs for the high and low index contrast systems as closely as possible, the high index contrast system contained 5 layers and the low contrast system contained 15 layers. Optimizations were performed for each system following the same methods described in the main text.

\begin{figure}[h]
  \centering
  \includegraphics[width=1.0\textwidth]{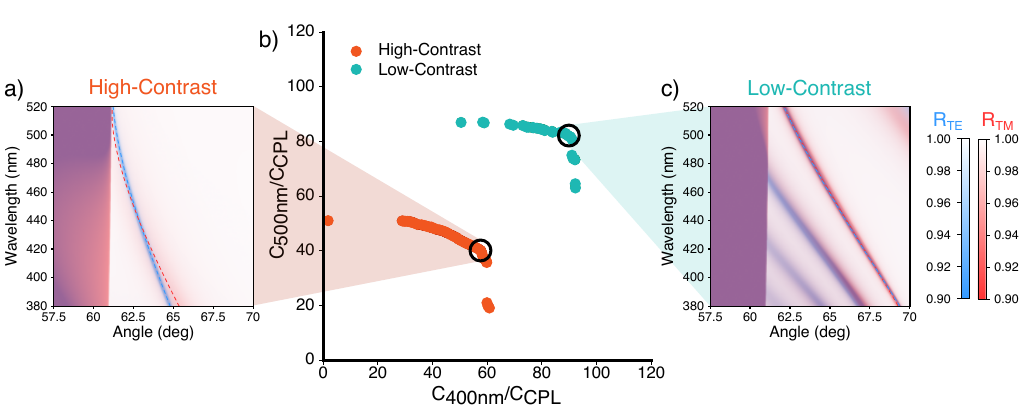}
  \caption{Results from an optimization with materials which display dispersion across the bandwidth of interest. Here the high index contrast designs are comprised of \ce{SiO2} and \ce{Ta2O5} while the low index contrast designs are comprised of Cellulose and Polystyrene.}
  \label{fig:real_mat_opt_results}
\end{figure}

The Pareto fronts from the multi-objective genetic optimization with systems using real materials are shown in figure \ref{fig:real_mat_opt_results} b). The reflectivity plots for the optimized design associated with the individual along the Pareto front with the highest combined evaluation function values, are shown in figure \ref{fig:real_mat_opt_results} a) and c) for the high and low index contrast designs respectively. The reflectivity plots show the same properties that were found in the results which did not account for material dispersion. Specifically we see that the high index contrast optimized design has a large anisotropy, as seen in the significantly wider modal width of the TM BSW as compared to the TE BSW, and more dissimilar dispersion relations for the TE and TM BSW modes as compared to the low index contrast system. These combined effects result in a lower optical chirality enhancement for the high index contrast system. Here, however, the low index contrast has a slightly narrower resonance than the high index contrast system making the comparison between the designs less fair as compared to the systems shown in the main manuscript. However, it is important to see the same behavior we saw in the case of the dispersionless materials while modeling real materials with dispersion.

\bibliographystyle{ieeetr}
\bibliography{LowcontrastBibtex}

\end{document}